# Lithium-ion-based solid electrolyte tuning of the carrier density in graphene

Jialin Zhao[1,2], Meng Wang[1,2], Hui Li[3], Xuefu Zhang[1,2], Lixing You[1,2], Shan Qiao[1,2], Bo Gao[1,2], Xiaoming Xie[1,2], Mianheng Jiang[1,2,4]

[1]State Key Laboratory of Functional Materials for Informatics, Shanghai Institute of Microsystem and Information Technology, Chinese Academy of Sciences, 865 Changning Road, Shanghai 200050, China.

[2]CAS Center for Excellence in Superconducting Electronics (CENSE), Shanghai 200050, China

[3]Shenzhen Key laboratory of Laser Engineering, College of Optoelectronic Engineering, Shenzhen University, Shenzhen, 518060, China

[4]School of Physical Science and Technology, ShanghaiTech University, 319 Yueyang Road, Shanghai, China.

**Abstract**

We have developed a technique to tune the carrier density in graphene using a lithium-ion-based solid electrolyte. We demonstrate that the solid electrolyte can be used as both a substrate to support graphene and a back gate. It can induce a change in the carrier density as large as $1\times10^{14}\,\mathrm{cm}^{-2}$, which is much larger than that induced with oxide-film dielectrics, and it is comparable with that induced by liquid electrolytes. Gate modulation of the carrier density is still visible at 150 K, which is lower than the glass transition temperature of most liquid gating electrolytes.

**Introduction**

The electronic properties of many materials strongly depend on their carrier density. In recent years, a technique called "liquid gating" has been widely used in materials research and development of flexible electronics[1-3]. It uses an ionic liquid or a polymer ion gel as a functional material to tune the carrier density in solid samples. The liquid electrolyte forms an electric double layer (EDL) at the solid-liquid interface, which consists of an image charge layer on the solid surface and a counterion layer in the liquid electrolyte. Because of the small separation between the two layers (0.3–0.5 nm), the EDL has a very large capacitance. It can induce a huge surface carrier density change of greater than $10^{14}$ cm$^{-2}$, which is at least two orders of magnitude greater than that induced with oxide-film dielectrics. The liquid gating technique has been successfully applied in the study of novel field-effect transistors[4-6] and superconductor–insulator transitions[7-11]. It has led to many exciting discoveries such as two-dimensional Ising superconductivity and gate modulation of the charge density wave[12-17]. However, the liquid gating technique has some drawbacks. The liquid nature of the electrolyte means that it is not compatible with most existing surface-probe techniques. Many liquid electrolytes are sensitive to the humidity in the environment, which degrades the performance of the liquid electrolyte. The sample in direct contact with the liquid electrolyte can also be damaged by the stress from the frozen electrolyte at low temperatures. Solid electrolytes can be used to overcome these drawbacks. It has been reported that a microporous SiO$_2$ thin film can be used to make a proton gate[18,19]. However，the proton-gating technique is very sensitive to the humidity in the environment. In addition, its performance in tuning the carrier density is still inferior to that of ionic liquids/polymer electrolytes. Fortunately a wide range of solid electrolytes were developed for battery research[20-22] and some of them can be suitable for the gating application. In this report, we demonstrate the potential of the lithium-ion-based solid electrolyte gating technique using single/few-layer graphene as a model system. We show that a lithium-ion conductive glass-ceramic (LICGC) can be used as both a substrate to support the graphene sample and a back gate to tune the carrier density. We demonstrate that the LICGC solid electrolyte can induce a change in the carrier density as large as $1 \times 10^{14}$ cm$^{-2}$, which is much larger than that induced by oxide dielectrics and comparable with that induced by liquid electrolytes. Additionally, the LICGC solid electrolyte can tune the carrier density in graphene samples at temperatures as low as 150 K, which is much lower than the glass transition temperature of most ionic liquid/polymer electrolytes.

**Sample preparation**

Single-layer graphene samples were initially grown on copper substrates by the chemical vapour deposition (CVD) technique. The growth details can be found in our previous paper[23]. The LICGC substrates (OHARA Inc., Sagamihara-shi, Japan), were cut into 5mm pieces using a dicing saw. Before the transfer of graphene, the LICGC substrates were rinsed in acetone and isopropanol, and then blow-dried with nitrogen. The single-layer graphene samples was transferred onto the LICGC substrates using the wet transfer method[24]. Figure 1a shows the steps of the wet transfer process. The typical size of the single-layer graphene samples was 3 mm × 4 mm. A 250-nm-thick Poly(methyl methacrylate) (PMMA) layer was first spin-coated on top of the graphene. The graphene/PMMA layer was detached from the copper substrate by using 10 wt. % $FeCl_3$ solution in deionized water. The floating graphene/PMMA layer was then transferred onto the LICGC substrate, and the PMMA layer was removed with acetone. The samples were then annealed in a hydrogen–argon atmosphere for 7 h. The electrical contacts to graphene were made from indium pellets. The few-layer graphene samples were mechanically exfoliated from highly oriented pyrolytic graphite onto the LICGC substrate using the well-known tape method. The exfoliated few-layer graphene were typically tens of micrometers in size. The standard e-beam lithography technique was used to pattern Ti/Au (5 nm/55 nm) microelectrodes. Figure 1b shows the steps of the patterning. The gate voltage was applied to the backside of the LICGC substrates using conductive silver paint.

**Results**

**Surface roughness of the LICGC substrate**

We used an atomic force microscope (AFM) to investigate the surface roughness of the LICGC substrate. Figure 2a shows an AFM topography image of a randomly selected area on the LICGC substrate. The measured average roughness is approximately 1.1 nm, which is similar to the surface roughness of the usual Si/SiO$_2$ substrate. The geometrical capacitance of the EDL is given by $C_g = \varepsilon_{se}\varepsilon_0/d$, where $\varepsilon_{se}$, $\varepsilon_0$, and $d$ are the relative permittivity of the solid electrolyte, the vacuum permittivity, and the spacing of the EDL. Because $C_g$ is inversely proportional to the spacing between the double layer, a large surface roughness may increase the local spacing of the EDL and thus reduce its geometrical capacitance. As we will show below, although not atomically flat, the LICGC substrate can create a considerable gating effect. Solid electrolytes with smaller surface roughness are always preferred. A solid electrolyte with an atomically flat surface is ideal to form a homogenous EDL, and also crucial for direct thin-film growth on solid electrolyte substrates using advanced thin-film growth technique such as molecular beam epitaxy (MBE), because it can give a clean and lattice-matched (with a right choice to material to grow) interface between the solid electrolyte substrate and the thin-film material..

**Tuning the carrier density in single/few-layer graphene**

We first demonstrate the gating effect of the solid electrolyte on single-layer graphene. Figure 2b shows the variation of the graphene sample resistance when sweeping the back gate voltage at room temperature. The inset shows a schematic diagram of the device geometry. The tuning of the carrier density can be considered as the charging/discharging process of a RC circuit. A double layer capacitor forms at the graphene/LICGC interface. $R_{Gr}$ and $R_{SE}$ are the resistance of the graphene sample and the LICGC substrate. $R_i$ is the interfacial resistance between the LICGC substrate and the silver paint, estimated to be greater than hundreds of megohms. By contrast, the resistance between the gold electrode and the liquid electrolyte, estimated from our own experiences, is in the megohm range. The ionic conductivity of LICGC is approximately 0.1 mS cm$^{-1}$ at room temperature, which is also one order of magnitude lower than that of DEME-BF$_4$ (a widely used ionic liquid)[25]. The large resistance in the RC circuit will lead to a long time constant. To minimize the gate sweep hysteresis, we chose to sweep the back gate voltage at a relatively slow speed (0.05–0.2 mV s$^{-1}$). As shown in Fig. 2b, the two resistance peaks in positive and negative gate sweeping are close to each other, suggesting a weak gate sweep hysteresis at room temperature.

To quantize the gate tuning capability of the solid electrolyte, we measured the carrier density change for a series of single- and few-layer graphene samples. The single-layer graphene covered a large portion of the LICGC substrate. The measurement for these samples gave an average gating effect of the solid electrolyte. The few-layer graphene samples had sizes on the micrometer scale. The gating effect on these small samples was more affected by the local surface roughness of the LICGC substrates. In both situations, we found that the LICGC substrate can realize considerable carrier density tuning. Figure 3a shows the variation of the resistance of a typical single-layer graphene sample with the back gate voltage. Figure 3b and 3c show the Hall coefficient and carrier density of the same sample measured at specific back gate voltages. Figure 3d–3f show similar measurements of a micro-fabricated few-layer graphene sample. The small resistance discontinuity in Fig. 3a and 3d is caused by the delayed (de)accumulation of Li ions at the LICGC/graphene interface due to the long RC time constant. We performed Hall measurements at each specific back gate voltage. The Hall measurements usually lasted 2h, during which time the electrical double layer capacitor was still in the process of charging or discharging, which led to discontinuity in the resistance ($R$) versus gate voltage ($V_g$) curves. Similar resistance discontinuity has been observed in liquid gating experiments[15], and more detailed experimental evidence of the delayed (de)accumulation of Li ions is given in the Supplementary Information. The single and few-layer graphene samples all showed clear bipolar behaviour. The sign changes of the Hall coefficients indicate that the Fermi level was tuned across the Dirac point. We observed a net electron injection (compared with the value at zero gate voltage) of approximately $5.0 \times 10^{13}$ cm$^{-2}$ in the wet transferred single-layer graphene sample, and approximately $1.0 \times 10^{14}$ cm$^{-2}$ in the exfoliated few-layer graphene sample. The observed gate tuning capability of LICGC on graphene is similar to the reported value for EMIM-BF4 ($1.0 \times 10^{14}$ cm$^{-2}$), and slightly smaller than that reported for DEME-TFSI ($2.5 \times 10^{14}$ cm$^{-2}$); which are both commonly used ionic liquid gating electrolytes[26].

**Temperature dependence of the performance of the LICGC solid electrolyte**

We investigated how the performance of the LICGC solid electrolyte varies with temperature. All liquid electrolytes have a specific glass-transition temperature, below which ion movement in the electrolytes is frozen and the electrostatic gate tuning ability is lost. To make a comparison with the liquid gating technique, we carried out back gate voltage sweeping experiments at various temperatures. Figure 4 shows the $R$ versus $V_g$ curves for five macroscopic single-layer graphene samples. These samples were cut from neighbouring areas on a large graphene sheet grown on the copper substrate. In these samples, the initial positions of the Dirac point should be similar. Gate modulation of the resistance is visible in all of the samples except for the one measured at 100K. This suggests that the Li ions are still displaceable at a temperature as low as 150 K, which is lower than the glass transition temperature of most known ionic liquids. We also found that the resistance peak shifted towards higher gate voltage as the temperature decreased, and the width of the resistance peaks became wider. This behaviour can be attributed to two possible reasons. First, because the gate tuning of the carrier density in graphene can be viewed as a charging/discharging process of the EDL capacitor, the RC time constant determines how fast the carrier density change occurs. The ionic conductivity of solid electrolytes follows Arrhenius law and decayed exponentially with temperature, which leads to increased resistance of the solid electrolyte. The interfacial resistance between LICGC substrate and the silver paint can also become greater at low temperature. The longer RC time constant leads to slower carrier density modulation. Second, at low temperatures, a high gate voltage is required to overcome the energy barrier for Li ions to approach the LICGC–graphene interface. Assuming that only the RC time constant affects the gate tuning behaviour, we expect that all of the resistance peaks should appear below a common voltage threshold after a sufficiently long time. However, we found the opposite case in the experiments: at low temperature, the resistance peak did not appear even after a long time unless we moved to a higher gate voltage. We thus believe that the combination of both effects results in the shift and widening of the graphene resistance peak. Detailed experimental evidence is given in the Supplementary Information.

**Discussion**

The results described above prove the concept of using solid electrolytes to tune the carrier density in thin-film samples. Because solid electrolytes are widely used in the research of Li-ion batteries and fuel cells, there is a long list of available materials that can be used to perform electrostatic gating. In the following paragraph, we will discuss the criteria for selecting a solid electrolyte for electrostatic gating.

It is essential to reduce the serial resistance in the RC circuit model to get a smaller RC time constant so that the carrier density change in the sample can keep up with the applied gate voltage. The interfacial resistance between the solid electrolyte and the gate electrode should be reduced, and the solid electrolytes with higher ionic conductivity are highly desired. The interfacial resistance is an active research field in Li-ion battery, which depends strongly on the material choices of solid electrolytes and cathode/anode electrodes[27]. The detailed discussion of the interfacial resistance is beyond the scope of this paper. In solid electrolytes, the ionic conductivity originates from defects, such as lattice vacancies and dopants occupying interstitial sites. Ions can hop between neighbouring defects. The ionic conductivity $\sigma$ can be expressed as $\sigma = nZ_e\mu$, where $n$ is the number of ions per unit volume, $Z_e$ is the charge of the ions, and $\mu$ is the ion mobility. As previously mentioned, the mobility follows the Arrhenius law and it exponentially decreases with decreasing temperature. Most solid electrolytes used in Li-ion battery research are good candidates for room temperature electrostatic gating, while the solid oxide electrolytes used in fuel cells usually only work at high temperatures. The LICGC substrate used in this study has an ionic conductivity of the order of $10^{-4}$ S cm$^{-1}$ at room temperature. Other electrolytes containing Li ions can reach of the order of $10^{-3}$ S cm$^{-1}$. By contrast, the ionic conductivity of organic liquid electrolytes is generally in the order of $10^{-2}$ S cm$^{-1}$. Materials with higher ionic conductivity are always favorable. Recently, new lithium solid electrolytes with high ionic conductivities have been reported, such as Li$_{10}$GeP$_2$S$_{12}$, which exhibits ionic conductivity higher than $10^{-2}$ S cm$^{-1}$ at room temperature. These materials are worthy of further investigation.

The second factor that needs to be taken into account is the surface roughness. As solid electrolytes are used as the substrate to support thin-film samples, a flat substrate surface is required. LICGC is one of the few commercially available solid electrolytes with a polished surface. To further improve this technique, a solid electrolyte with an atomically flat surface is highly desired. Such solid electrolytes may form a homogeneous EDL with large geometrical capacitance, which can enhance the electrostatic gating effect. Furthermore, a solid electrolyte with an atomically flat surface is an ideal substrate for advanced thin-film growth techniques such as MBE. It can create a much cleaner interface between the solid electrolyte and the thin-film sample compared with the interface made by transferring thin-film samples to the surface of a solid electrolyte. Although most of the solid electrolytes developed for energy storage research are in ceramic or polycrystalline forms, there are some single-crystal solid electrolytes, such as potassium-doped β-alumina. This will be important to advance the solid electrolyte gating technique.

Some other factors should be considered. The chemical stability under ambient conditions is important for practical applications. LICGC is a good material in this sense because it is not sensitive to humidity, unlike many ionic liquids. Solid electrolytes with high electrochemical decomposition potentials are also preferable so that large gate voltages can be applied. Using LICGC as the back gate, the gate voltage is limited between −2 and 3.5 V. We also attempted to apply higher gate voltages. As shown in the Supplementary Information, a higher negative gate voltage led to an increase of the graphene sample resistance. We observed such behaviour in both single-layer graphene samples with indium pellet contacts and few-layer graphene samples with Ti–Au electrode contacts. In contrast, applying a high positive back gate voltage of up to 40 V did not cause a significant increase of the sample resistance. However, as shown in Fig. 3b, the electron density measured in other single- or few-layer graphene samples saturated at a high positive gate voltage. Similar behaviour has been reported for $Bi_2Te_3$ gated with an ionic liquid[5], and it was attributed to the limitation of the chemical window of the ionic liquid. The performance of the LICGC solid electrolyte at high positive/negative gate voltages is also probably related to the restriction of the electrochemical window. This is still under investigation.

In summary, we have discussed the advantages of the LICGC solid electrolyte gating technique over the existing liquid gating technique. The LICGC solid electrolyte can serve as both the substrate to support the sample and the back gate. It can achieve a similar carrier density tuning ability to liquid electrolytes, while maintaining an uncovered sample surface. Therefore, it provides the possibility of combining the electrostatic carrier density tuning technique with other surface-probe techniques. The combination of these techniques will be very useful, for example, to monitor evolution of the Fermi level by continuously tuning the carrier density in superconducting materials. LICGC substrates provide a strong and full solid-state gate tuning capability, making them more suitable for practical applications than liquid electrolytes. A sample fabricated on top of a LICGC substrate will also not suffer from problems that may be encountered in liquid gating, such as damage caused by the stress created during the freezing of the liquid electrolyte. All of these advantages make solid electrolyte gating a promising technique for future applications.

## References

1   Tucceri, R. A review about the surface resistance technique in electrochemistry. *Surf. Sci. Rep.* **56**, 85-157, doi: 10.1016/j.surfrep.2004.09.001 (2004).

2   Misra, R., McCarthy, M. & Hebard, A. F. Electric field gating with

ionic liquids. *Appl. Phys. Lett.* **90**, 052905, doi:10.1063/1.2437663 (2007).

3   Cho, J. H. *et al.* Printable ion-gel gate dielectrics for low-voltage polymer thin-film transistors on plastic. *Nat. Mater.* **7**, 900-906, doi:10.1038/nmat2291 (2008).

4   Kim, B. J. *et al.* High-Performance Flexible Graphene Field Effect Transistors with Ion Gel Gate Dielectrics. *Nano Lett.* **10**, 3464-3466, doi:10.1021/nl101559n (2010).

5   Yuan, H. *et al.* Liquid-Gated Ambipolar Transport in Ultrathin Films of a Topological Insulator $Bi_2Te_3$. *Nano Lett.* **11**, 2601-2605, doi:10.1021/nl201561u (2011).

6   Pu, J. *et al.* Highly Flexible $MoS_2$ Thin-Film Transistors with Ion Gel Dielectrics. *Nano Lett.* **12**, 4013-4017, doi:10.1021/nl301335q (2012).

7   Ye, J. T. *et al.* Liquid-gated interface superconductivity on an atomically flat film. *Nat. Mater.* **9**, 125-128, doi:10.1038/nmat2587 (2010).

8   Ueno, K. *et al.* Discovery of superconductivity in $KTaO_3$ by electrostatic carrier doping. *Nano Lett.* **6**, 408-412, doi:10.1038/nnano.2011.78 (2011).

9   Ye, J. T. *et al.* Superconducting Dome in a Gate-Tuned Band Insulator. *Science* **338**, 1193-1196, doi:10.1126/science.1228006

(2012).

10   Saito, Y., Kasahara, Y., Ye, J., Iwasa, Y. & Nojima, T. Metallic ground state in an ion-gated two-dimensional superconductor. *Science* **350**, 409-413, doi:10.1126/science.1259440 (2015).

11   Shi, W. *et al.* Superconductivity Series in Transition Metal Dichalcogenides by Ionic Gating. *Sci. Rep.* **5**, 12534, doi:10.1038/srep12534 (2015).

12   Efetov, D. K. & Kim, P. Controlling Electron-Phonon Interactions in Graphene at Ultrahigh Carrier Densities. *Phys. Rev. Lett.* **105**, 256805, doi:10.1103/PhysRevLett.105.256805 (2010) (2010).

13   Lee, M., Williams, J. R., Zhang, S., Frisbie, C. D. & Goldhaber-Gordon, D. Gate-Controlled Kondo Effect in $SrTiO_3$. *Phys. Rev. Lett.* **107**, 256601, doi:10.1103/PhysRevLett.107.256601 (2011).

14   Daghero, D. *et al.* Large Conductance Modulation of Gold Thin Films by Huge Charge Injection via Electrochemical Gating. *Phys. Rev. Lett.* **108**, 066807, doi:10.1103/PhysRevLett.108.066807 (2012).

15   Li, Z. J., Gao, B. F., Zhao, J. L., Xie, X. M. & Jiang, M. H. Effect of electrolyte gating on the superconducting properties of thin $2H-NbSe_2$ platelets. *Supercond. Sci. Tech* **27**, 015004, doi:10.1088/0953-2048/27/1/015004 (2014).

16  Lu, J. M. *et al.* Evidence for two-dimensional Ising superconductivity in gated MoS$_2$. *Science* **350**, 1353-1357, doi:10.1126/science.aab2277 (2015).

17  Yu, Y. *et al.* Gate-tunable phase transitions in thin flakes of 1T-TaS$_2$. *Nat. Nanotechnol.* **10**, 270-276, doi:10.1038/nnano.2014.323 (2015).

18  Lu, A., Sun, J., Jiang, J. & Wan, Q. Microporous SiO$_2$ with huge electric-double-layer capacitance for low-voltage indium tin oxide thin-film transistors. *Appl. Phys. Lett.* **95**, 222905, doi:10.1063/1.3271029 (2009).

19  Chao, J. Y., Zhu, L. Q., Xiao, H. & Yuan, Z. G. Protonic/electronic hybrid oxide transistor gated by chitosan and its full-swing low voltage inverter applications. *J. Appl. Phys.* **118**, 235301, doi:10.1063/1.4937555 (2015).

20  Kamaya, N. *et al.* A lithium superionic conductor. *Nat. Mater.* **10**, 682-686, doi:10.1038/nmat3066 (2011).

21  Li, M. *et al.* A family of oxide ion conductors based on the ferroelectric perovskite Na$_{0.5}$Bi$_{0.5}$TiO$_3$. *Nat. Mater.* **13**, 31-35, doi:10.1038/nmat3782 (2014).

22  Wang, Y. *et al.* Design principles for solid-state lithium superionic conductors. *Nat. Mater.* **14**, 1026-1031, doi:10.1038/nmat4369 (2015).

23  Wu, T. *et al*. Fast growth of inch-sized single-crystalline graphene from a controlled single nucleus on Cu-Ni alloys. *Nat. Mater.* 15, 43-47, doi:10.1038/nmat4477(2016)

24  Suk, J. W. *et al.* Transfer of CVD-Grown Monolayer Graphene onto Arbitrary Substrates. *ACS Nano* **5**, 6916-6924, doi:10.1021/nn201207c (2011).

25  Sato, T., Masuda, G. & Takagi, K. Electrochemical properties of novel ionic liquids for electric double layer capacitor applications. *Electrochim. Acta* **49**, 3603-3611, doi:10.1016/j.electacta.2004.03.030 (2004).

26  Ye, J. *et al.* Accessing the transport properties of graphene and its multilayers at high carrier density. *Proc. Natl. Acad. Sci.* **108**, 13002-13006, doi:10.1073/pnas.1018388108 (2011).

27  Wu, B. *et al*. Interfacial Behaviours Between Lithium Ion Conductors and Electrode Materials in Various Battery Systems. *J. Mater. Chem. A*, doi:10.1039/C6TA05439K (2016).

**Acknowledgement**

We acknowledge that the research was inspired by Prof. Xianhui Chen's lecture in last Novenmber, in which Prof. Chen presented his application of solid electrolyte gating technique in the study of FeSe superconductivity, and that one of the contributing authors discussed with Prof. Chen about solid electrolyte gating after the lecture. We thank Chilin Li for providing the information of the solid electrolytes and Xiaoyu Liu for the help with nano-pattering. We acknowledge the support from the "Strategic Priority Research Program (B)" of the Chinese Academy of Sciences under Grant No. XDB04010600 and No. XDB04030000; from the National Natural Science Foundation of China under Grant No. 11374321; and from Helmholtz Association through the Virtual Institute for Topological Insulators (VITI).

**Author Information**

**Affiliations**

**State Key Laboratory of Functional Materials for Informatics, Shanghai Institute of Microsystem and Information Technology, Chinese Academy of Sciences, 865 Changning Road, Shanghai 200050, China.**

**CAS Center for Excellence in Superconducting Electronics (CENSE), Shanghai 200050, China**

Jialin Zhao, Meng Wang, Xuefu Zhang, Lixing You, Shan Qiao, Bo GAO, Xiaoming Xie, Mianheng Jiang

**Shenzhen Key laboratory of Laser Engineering, College of Optoelectronic Engineering, Shenzhen University, Shenzhen, 518060, China**

Hui Li

**School of Physical Science and Technology, ShanghaiTech University, 319 Yueyang Road, Shanghai, China.**

Mianheng Jiang

**Contributions**

J. Zhao, M.Wang and H. Li performed the experiments. X. Zhang provided single-layer graphene sheets. L. You, S. Qiao, X. Xie and M. Jiang contributed to the discussions. B. Gao designed the experiment and wrote the manuscript.

## Competing interests

The authors declare no competing financial interests.

## Corresponding Authors

Correspondence to Bo GAO (bo_f_gao@mail.sim.ac.cn)

**Figure legends:**

**Figure 1**: (a) Schematic diagram of the wet transfer of single-layer graphene from Cu substrate to LICGC substrate. (b) Schematic diagram of patterning metal electrodes on few-layer graphene exfoliated onto the LICGC substrate.

**Figure 2**: (a) The AFM image of a LICGC substrate, the average roughness is approximately 1.1 nm. (b) The bipolar behavior of a typical single-layer graphene sample measured at room temperature. Inset: schematic view of the device geometry.

**Figure 3**: The resistance (a), Hall coefficient (b) and carrier density (c) of a single-layer graphene sample as a function of the back gate voltage measured at room temperature. (d) – (f) similar measurement results of a micrometer-sized few-layer graphene sample.

**Figure 4**: The gate dependence of five single-layer graphene samples; each measured at a specific temperature. Li ions are displaceable even at a temperature as low as 150K.

**Figure 1**

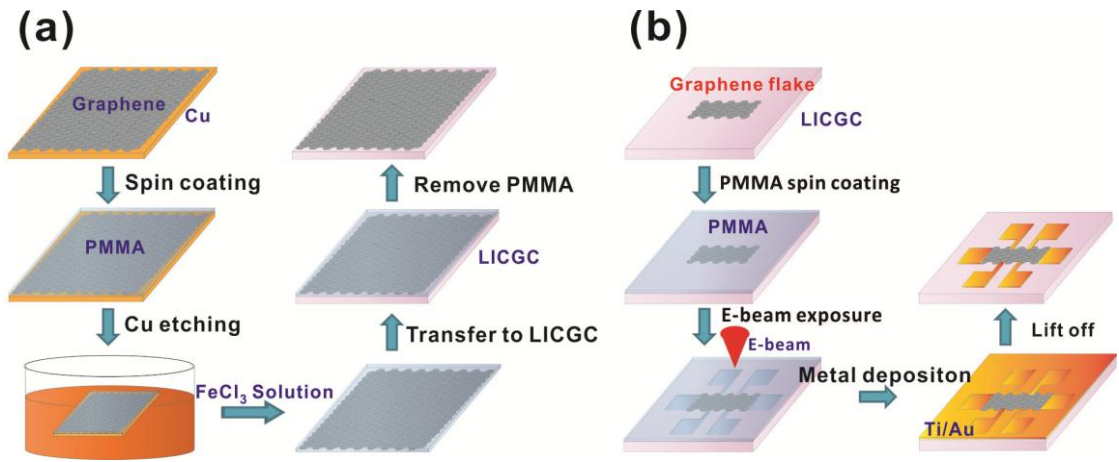

**Figure 2**

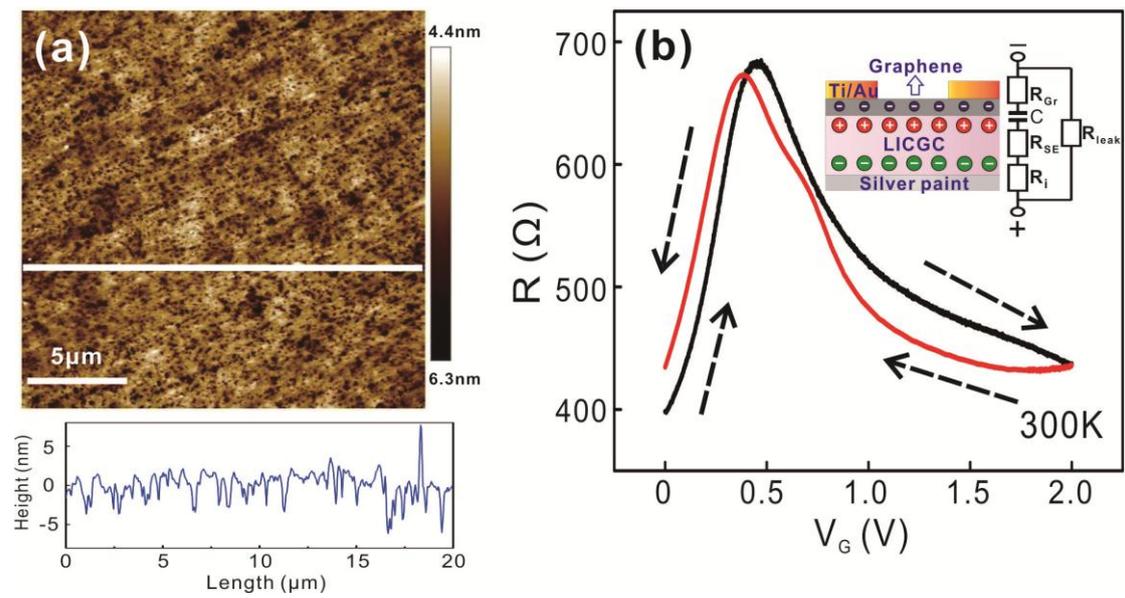

**Figure 3**

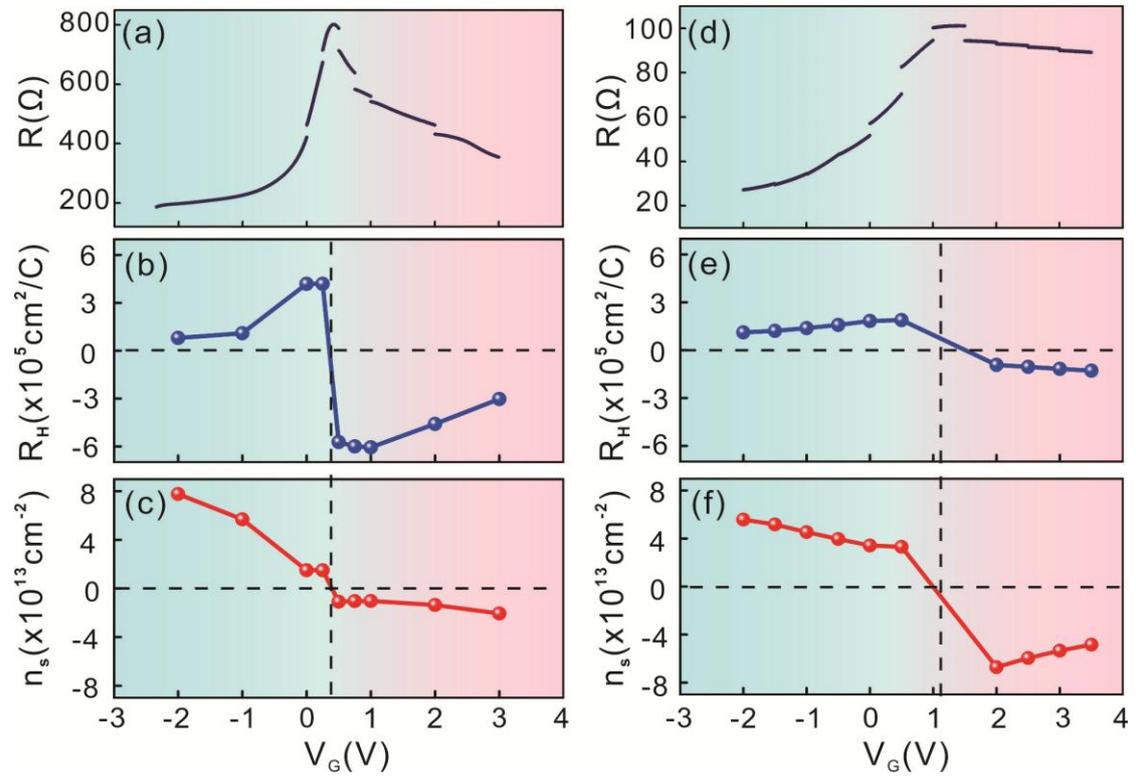

**Figure 4**

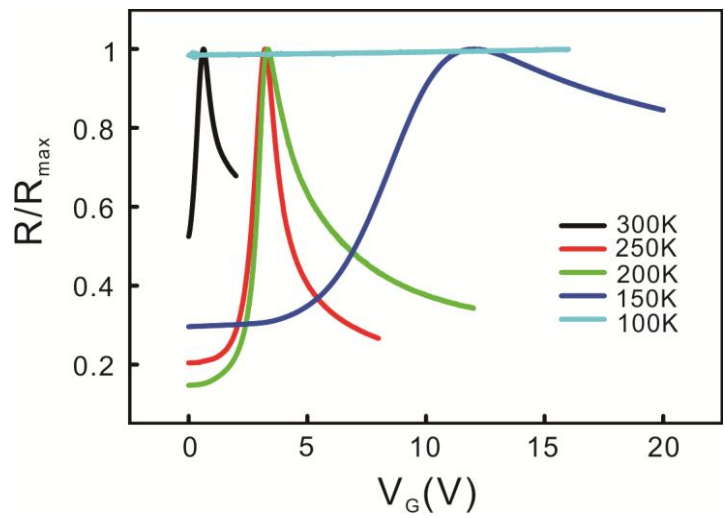

# Supplementary Materials for

# Lithium-ion-based solid electrolyte tuning of the carrier density in graphene

Jialin Zhao[1,2], Meng Wang[1,2], Hui Li[3], Xuefu Zhang[1,2], Lixing You[1,2], Shan Qiao[1,2], Bo Gao[1,2], Xiaoming Xie[1,2], Mianheng Jiang[1,2,4]

[1]State Key Laboratory of Functional Materials for Informatics, Shanghai Institute of Microsystem and Information Technology, Chinese Academy of Sciences, 865 Changning Road, Shanghai 200050, China.

[2]CAS Center for Excellence in Superconducting Electronics (CENSE), Shanghai 200050, China

[3]Shenzhen Key laboratory of Laser Engineering, College of Optoelectronic Engineering, Shenzhen University, Shenzhen, 518060, China

[4]School of Physical Science and Technology, ShanghaiTech University, 319 Yueyang Road, Shanghai, China.

E-mail: bo_f_gao@mail.sim.ac.cn

**This file includes:**

Fig. S1 and Fig. S2

Fig. S1

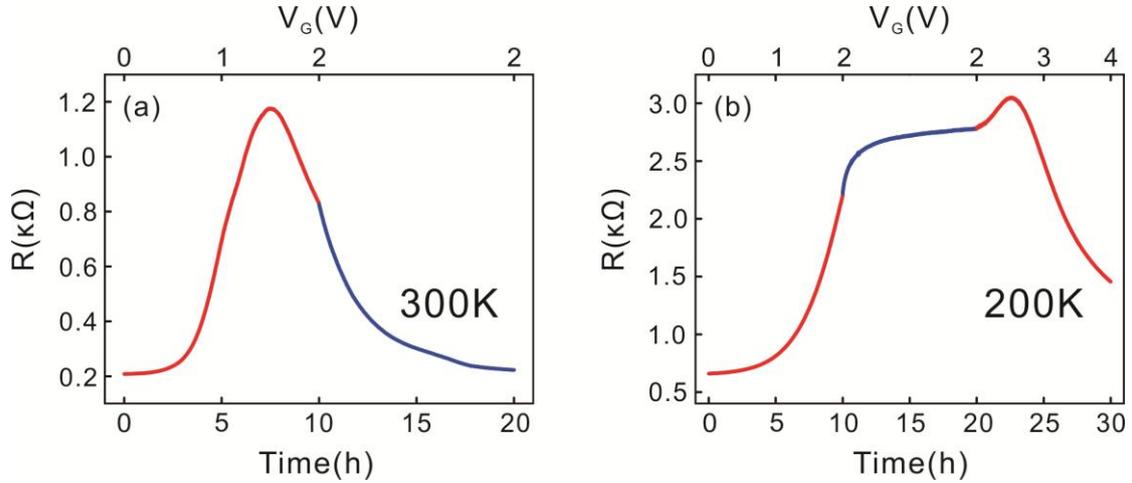

We carried out two experiments to demonstrate the evolution of Li-ions accumulation at the LICGC–graphene interface and the decline of the performance of the LICGC substrate in tuning the carrier density at low temperature. We used newly grown single-layer graphene whose resistance peak appear near 1.5 V (gate voltage sweep rate ~ 0.05 mV s$^{-1}$) at room temperature. Figure S1a shows the evolution the graphene resistance with time and applied gate voltage at room temperature. We swept the gate voltage from 0 to 2 V, as shown by the curve in red. We found that the resistance peak appeared near 1.5 V. We kept the back gate voltage at 2 V and recorded the change of the sample resistance with time, shown by the curve in blue. The resistance of the sample continued to drop, suggesting that the double layer capacitor was being charged. This is due to the long RC time constant. It explains the resistance discontinuity in Fig. 3a and 3d in the main text. Figure S1b shows similar measurements performed at 200 K. The gate voltage was first swept from 0 to 2 V. The resistance peak did not appear near 1.5V. If the only change at low temperature had been the enlarged RC time constant, we would expect that the resistance peak appear after a sufficiently long time. However, the resistance saturated after 10 hours. The resistance peak finally appeared after we swept the gate voltage to 3 V. This experiment suggests that the LICGC substrate shows weaker performance when used to tune the carrier density at low temperature. It loses completely the ability to tune the carrier density below 100 K, as shown in Fig. 4 in the main text. The possible explanation is that a higher back gate voltage may be required to overcome the energy barrier encountered by lithium ions moving toward the LICGC–graphene interface at low temperature.

Fig. S2

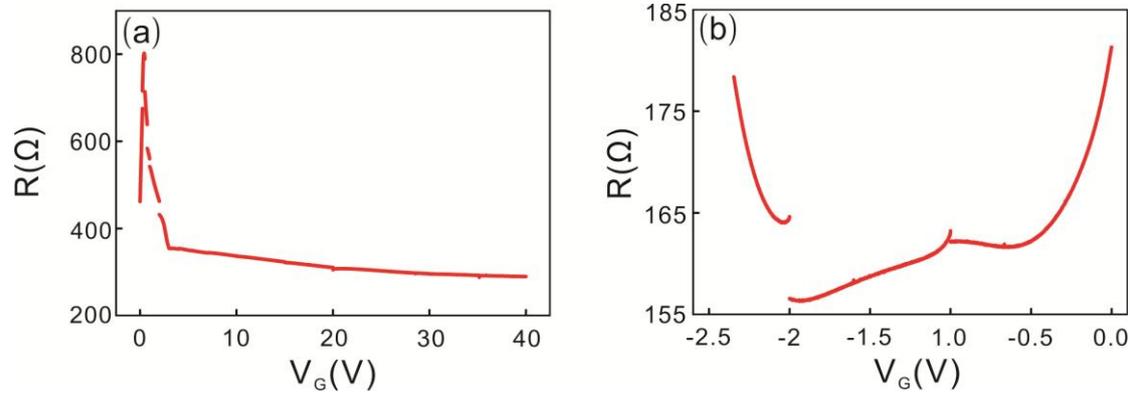

We also tried to apply high back gate voltage to investigate the performance of the LICGC solid electrolyte in such condition. We found that high positive gate voltage did not degrade significantly the graphene sample, while high negative gate voltage caused an abrupt rising of the sample resistance. The discontinuity in the above curves is due to the delayed Li ion displacements during the time interval in the measurements. It is known that electro-chemical reaction may take place when gating with liquid electrolytes. And sometimes it may bring additional electro-chemical doping effect that enhances the gate tuning effect [1,2]. We are investigating if similar electro-chemical doping also take places in the solid electrolyte gating process.

1 Li, Z. J., Gao, B. F., Zhao, J. L., Xie, X. M. & Jiang, M. H. Effect of electrolyte gating on the superconducting properties of thin 2H-NbSe$_2$ platelets. *Supercond. Sci. Tech* **27**, 015004, doi:10.1088/0953-2048/27/1/015004 (2014)..

2 Yu, Y. *et al.* Gate-tunable phase transitions in thin flakes of 1T-TaS$_2$.

*Nat. Nanotechnol.* **10**, 270-276, doi:10.1038/nnano.2014.323 (2015).